\documentstyle[12pt]{article}

\newcommand{\be}{\begin{equation}}
\newcommand{\ee}{\end{equation}}
\newcommand{\bea}{\begin{eqnarray}}
\newcommand{\eea}{\end{eqnarray}}

\newcommand{\I} {{\cal I}}

\begin{document}

\begin{center}
\begin{large}
{\bf  Radiation via Tunneling \\}
{\bf from a \\}
{\bf de Sitter Cosmological Horizon \\}
\end{large}  
\end{center}
\vspace*{0.50cm}
\begin{center}
{\sl by\\}
\vspace*{1.00cm}
{\bf A.J.M. Medved\\}
\vspace*{1.00cm}
{\sl
Department of Physics and Theoretical Physics Institute\\
University of Alberta\\
Edmonton, Canada T6G-2J1\\
{[e-mail: amedved@phys.ualberta.ca]}}\\
\end{center}
\bigskip\noindent
\begin{center}
\begin{large}
{\bf
ABSTRACT
}
\end{large}
\end{center}
\vspace*{0.50cm}
\par
\noindent
 
Hawking radiation can usefully be viewed as
a  semi-classical tunneling process that originates at the black hole horizon. 
The same basic premise should apply to
de Sitter background radiation,  with
the  cosmological horizon of de Sitter space now playing
 the  featured role. 
In fact, a recent work [hep-th/0204107] 
has gone a long way to verifying the validity of this de Sitter-tunneling
picture.
 In the current paper, we extend these
prior considerations  to arbitrary-dimensional de Sitter space,
 as well as Schwarzschild-de Sitter spacetimes.
It is shown that the tunneling formalism
naturally  censors against any black
hole with a mass in excess of the Nariai value;
thus enforcing  a ``third law'' of Schwarzschild-de Sitter thermodynamics.
We also provide commentary  on the dS/CFT correspondence
in the context of this tunneling framework.

\newpage

\section{Introduction}
\par

In light of recent astronomical observations,  it
has been suggested  that  our universe  
will  asymptotically approach a de Sitter spacetime \cite{BOPS}.
This realization has sparked a sense of urgency in
resolving the quantum-gravitational mysteries
of de Sitter space \cite{WIT2}. With prompting from
the  very successful  anti-de Sitter/conformal
field theory correspondence \cite{MAL,GUB,WIT},
much of this work has focused on finding a  
holographic description \cite{THO,SUS}  of de Sitter space.
In particular, there has been much ado about 
establishing an analogous duality; that is,  the
so-called dS/CFT correspondence \cite{STR}. (Also see, for instance,
\cite{KLE}-\cite{SV}.) Although there has
been considerable success along this line, the proposed
duality is still marred by various ambiguities.
For example, a dual boundary theory that appears to
be a non-unitary one \cite{STR},  a conspicuous
absence of measurable quantities (at least those
with operational meaning to a de Sitter observer \cite{WIT2}),
and  logistic breakdowns \cite{SUS2} that can be attributed
to the finite entropy of de Sitter space \cite{BOU}.
\par
The above issues should probably  be viewed as
significant philosophical roadblocks as opposed to
mere technical difficulties. Which is to say,
their resolution will, in all likelihood,
necessitate dramatic departures from  our
current ways of thinking. Hence, it may be an appropriate
juncture to ``take a step back'' and re-enforce our
understanding of de Sitter space at a semi-classical level.
With  this as our mind-set, let us now proceed to consider
the central topic of the paper; namely, de Sitter radiation
as a semi-classical tunneling process.
\par
First, let us briefly review the concept of  tunneling  
as it applies to  a radiating black hole.  According to
this interpretation, Hawking radiation \cite{HAW}
can be attributed to the spontaneous creation
of particles at a point just inside of
the black hole horizon. One of the particles then tunnels
out to the opposite side of the horizon, where
it emerges with positive energy. Meanwhile, the
negative-energy ``partner''  remains behind and effectively
lowers the mass of the black hole. 
\par
The above point of view formed the foundation 
for a program of study  that was initiated  by Kraus
and Wilczek (KW) \cite{KW}\footnote{For further
developments and generalizations, see
\cite{KW2}-\cite{PAR}. 
For other perspectives on radiation via tunneling, see \cite{MAPA,SSP}.}
and is central to the current work.
The essence of the KW  methodology is a dynamical treatment
of black hole radiation. More to the point, KW considered
the effects of a self-gravitating matter shell propagating
outwards through a spherically symmetric black hole
horizon. Two particularly significant points of
this work are as follows. {\it (i)} The background geometry
is allowed to fluctuate so that the formalism incorporates
a black hole of varying mass. In this manner, the total
energy of the spacetime is naturally conserved. Notably, 
energy conservation is often overlooked in other formal
treatments of Hawking radiation \cite{BD}.
{\it (ii)} Boundary conditions are imposed by foliating
the spacetime with somewhat unconventional
``Painleve coordinates'' \cite{PAIN}. Significantly, 
these coordinates are regular at the horizon, as
well as stationary but {\it} not static (i.e., time-reversal
asymmetry is manifest). This gauge seems quite appropriate
for describing the geometry of a slowly-evaporating black
hole. 
\par
Let us now return the discussion to de Sitter space. As it is well known,
there are many similarities between the thermodynamic properties of 
a de Sitter cosmological horizon and those of a black hole
horizon \cite{GH}.
Hence, it would seem natural to extend the tunneling picture
and KW treatment to the background radiation associated
with de Sitter space.  Just such a study was
recently carried out by Parikh \cite{PAR} with
considerable success. This author, however, focused on
the interesting but unphysical case of
3-dimensional de Sitter space. The main purpose of the
current paper is to generalize considerations to 
a de Sitter spacetime of arbitrary dimensionality.
We will also  provide some commentary on the
dS/CFT correspondence in the context of this study.
\par
The remainder of the paper is organized as follows.
In Section 2, we consider
a radiating cosmological horizon in an empty, $n$+2-dimensional
de Sitter spacetime and,
 with guidance from \cite{KW,PW,PAR},
 calculate the semi-classical emission rate.
The consistency of the derived expression is
then verified  for the case of $n=3$.  We
 accomplish this by extrapolating the emission spectrum
and comparing the lowest 
order term  with  standard de Sitter thermodynamics.
In Section 3, we further extend the formalism
to a Schwarzschild-de Sitter spacetime. Here, we also
comment on thermal stability and touch upon the subject of 
 dS/CFT  renormalization group flows \cite{STR2}.
In Section 4, we take a  step towards the ethereal
and  reconsider the tunneling picture  
from an outside-of-the-horizon perspective.
The dS/CFT correspondence  provides the
motivating factor for this  portion of the analysis.
Finally, Section 5 contains a brief summary.

\section{De Sitter Tunneling}
\par
Let us begin by considering an $n$+2-dimensional de Sitter
spacetime (with $n\geq 1$). There are many different coordinate
systems that can be used to provide a local  description
of de Sitter space \cite{SSV},
including the following explicitly static coordinates:
\be
ds^2=-\left(1-{r^2\over l^2}\right)dt^2+
      \left(1-{r^2\over l^2}\right)^{-1}dr^2+r^2d\Omega_n^2.
\label{1}
\ee
Here, $l$ is the curvature radius of de Sitter
space (i.e., $\Lambda=n(n+1)/2l^2$ is the 
positive cosmological constant), $d\Omega^2_n$  represents
 an $n$-dimensional spherical hypersurface of unit radius, and
the non-angular coordinates range  according to $0\leq r\leq l$
and $-\infty \leq t \leq +\infty$.
Keep in mind that the boundary at  $r=l$ describes a cosmological
horizon for an observer located at $r=0$.
\par
The above coordinates  fail, of course, to cover the entire
de Sitter manifold.  Eq.(\ref{1}) does, however, precisely cover
the so-called ``southern causal diamond'' \cite{SSV}, which is 
the region of spacetime that is fully accessible to an observer
at the south pole ($r=0$). A particularly attractive
feature of this coordinate gauge is the existence of
a timelike Killing vector ($\partial_t$), thus leading to
a sensible notion of time evolution for a south-pole observer.
Note that such a timelike Killing vector is notoriously
absent in {\it any} global description of de Sitter space.
\par
As discussed in Section 1,
it is most convenient, in the tunneling picture, to use stationary 
coordinates  that are
 manifestly asymmetric under time reversal.  
In the case of a Schwarzschild black hole, the following
Painleve coordinates \cite{PAIN} have been utilized
for just this purpose \cite{KW}:
\be
ds^2=-\left(1-{2M\over r}\right)d\tau^2-2\sqrt{{2M\over r}}d\tau dr
+dr^2+r^2 d\Omega^2_2.
\label{2}
\ee
Along with the above-mentioned properties, this system
has the distinguishing  features of horizon  regularity  and
flat  constant-time surfaces.  
Further note that  such coordinates comply with the perspective
of a free-falling observer, who is expected to
experience nothing out of the ordinary upon passing through
the horizon.
\par
To obtain an analogous coordinate system for de Sitter space
or ``Painleve-de Sitter'' coordinates \cite{PAR},
we first employ the following transformation:
\be
t=\tau+f(r),
\label{3}
\ee
so that the static metric (\ref{1}) takes on the form:
\bea
ds^2=&-&\left(1-{r^2\over l^2}\right)d\tau^2 - 2{df\over dr}
\left(1-{r^2\over l^2}\right)d\tau dr \nonumber \\
&+& \left[\left(1-{r^2\over l^2}\right)^{-1}-
\left(1-{r^2\over l^2}\right)\left({df\over dr}\right)^2\right]
 dr^2+r^2 d\Omega^2_n.
\label{4}
\eea
\par
Next, we enforce that constant-$\tau$ slices reduce to
$n$+1-dimensional flat space  (i.e., $dr^2+r^2d\Omega^2_n$),
so that:
\be
{df\over dr}=\pm {r\over l} \left(1-{r^2\over l^2}\right)^{-1}.
\label{5}
\ee
Hence, we can rewrite the metric (\ref{4}) as:
\be
ds^2=-\left(1-{r^2\over l^2}\right)d\tau^2 - 2{r\over l}
d\tau dr 
+ dr^2+r^2 d\Omega^2_n,
\label{6}
\ee
up to an arbitrary choice of sign in the off-diagonal term.
\par
Clearly,  these ``new'' coordinates
exhibit all of the priorly discussed features of Painleve
coordinates; including horizon ($r=l$) regularity and
time-reversal asymmetry. The feature of horizon regularity
has special significance in de Sitter space, as the
revised coordinates are no longer restricted to
the southern causal diamond. In fact, Eq.(\ref{6})
covers the entire causal future (of an observer
at $r=0$), which translates to precisely one half of the
complete de Sitter manifold. Meanwhile, one can 
describe the remaining half (or the causal
past) by ``flipping'' the sign in front of the off-diagonal term.
In this sense, the Painleve-de Sitter coordinate system
is  closely related to de Sitter planar coordinates
\cite{SSV}; a point which has been elaborated on
in \cite{PAR}. Of further interest, $\partial_{\tau}$ 
is a Killing vector throughout the Painleve-de Sitter
system, although this vector changes character
(timelike $\rightarrow$ spacelike) upon passing through
the horizon.
\par
For later usage, let us  evaluate the radial, null geodesics
described by Eq.(\ref{6}). Under these conditions
($d\Omega^2_n=ds^2=0$), we can re-express this line element
as follows:
\be
({\dot r})^2 -2{r\over l}{\dot r}-\left(1-{r^2\over l^2}\right)
=0,
\label{7}
\ee
where a dot denotes differentiation with respect to $\tau$.
Solving the quadratic, we then have:
\be
{\dot r}= {r\over l}\pm 1,
\label{8}
\ee
where the  $+/-$ sign  can be identified with outgoing/incoming radial
motion.  
\par
With the de Sitter cosmological horizon (at $r=l$)
in mind, let us now focus on a semi-classical treatment of
the associated  radiation, as advocated in (for instance)  \cite{KW,PW,PAR}.
First of all, we adopt the picture of a pair of particles spontaneously
created just outside of the  horizon.  
The positive-energy particle tunnels through the horizon to
emerge as an inward-moving, self-gravitating   energy shell; whereas the 
negative-energy particle remains
behind and effectively lowers the energy of the background spacetime. 
Because of the infinite blue-shift near the horizon,
the emerging energy shell can be treated as a point particle;
meaning that a WKB-type of approximation may be appropriately
employed. For sake of simplicity, we will further
invoke an ``s-wave'' approximation;  in particular, we assume a massless
shell and  symmetry with respect to the angular coordinates.
\par
Given this semi-classical, WKB framework, it has been shown that the 
logarithm of the emission rate ($\Gamma$)
can be expressed in terms of the imaginary part
of the ``total'' (particle plus gravitational) action, $\I$ \cite{KW}.  
More specifically:
\be
\Gamma \approx e^{-2 Im\I}.
\label{9}
\ee
Alternatively, one can re-express this relation in the
following spectral form: 
\be
{\omega\over T(\omega)}\approx 2Im\I,
\label{10}
\ee
where $\omega>0$ is the  particle energy 
and $T(\omega)$ can be identified  with the effective
temperature.
\par
For a positive-energy s-wave, the imaginary part of the 
action has been found to have a conveniently simple
form \cite{KW}:
\be
Im\I =Im\int d\tau {\dot r}
 p_r =Im\int^{r_{f}}_{r_{i}}\int^{p_r}_{0}
dp^{\prime}_r dr,
\label{11}
\ee 
where $p_r$ is the canonical momentum (conjugate to $r$).
Also, $r_i$ and $r_f$ indicate (roughly) the point of particle
creation and the classical turning point of motion.
\par
To proceed with an explicit calculation, it is
useful to apply Hamilton's equation:
\be
{\dot r}={dH\over dp_r}={d(E-\omega)\over dp_r}=-{d\omega\over
dp_r}.
\label{12}
\ee
Here, $E$  represents the total conserved energy
of the system, whereas $E-\omega$  can be
regarded as the (varying) gravitational
energy stored in the background spacetime. 
Let us re-emphasize that, by keeping $E$ fixed,
energy conservation will be enforced in a natural
way.
\par
Substituting Eq.(\ref{12}) into Eq.(\ref{11}), we
find:
\be
Im\I=-Im\int^{r_f}_{r_i}\int^{\omega}_{0} {d\omega^{\prime}dr\over {\dot r}}.
\label{12.5}
\ee
\par
Before  evaluating  the above integral, we must 
necessarily obtain an expression for ${\dot r}$ as
a function of $\omega^{\prime}$. The form of this expression
depends on the answer to the following question: what
effective metric does the energy shell see as it propagates
through the background spacetime?  Considering that the de Sitter
background loses some of its energy to the propagating shell,
we propose that the effective metric in question is
that of a Schwarzschild-de Sitter geometry. The reasoning
is somewhat subtle and  based on the observation
that the total  energy of a Schwarzschild-de Sitter spacetime 
is always more negative than that of empty de Sitter space.
That is to say, the energy of a Schwarzschild-de Sitter
spacetime is known to decrease with 
 increasing black hole mass \cite{BDM}. Although counter-intuitive,
this inverted correspondence can be attributed
to a negative binding energy between a positive-mass object
and a de Sitter gravitational field \cite{MYU}.
\par
With the above discussion in mind, let us now consider
 Schwarzschild-de Sitter static coordinates:
\be
ds^2=-\left(1-{r^2\over l^2}-{M\epsilon_n\over r^{n-1}}\right)dt^2+
      \left(1-{r^2\over l^2}-{M\epsilon_n\over r^{n-1}}
\right)^{-1}dr^2+r^2d\Omega_n^2,
\label{13}
\ee
where
$\epsilon_{n}\equiv 16\pi G_{n+2}/n{\cal V}_n$ and
 $M$ is the conserved  mass \cite{BDM}. 
(Also, $G_{n+2}$ is  Newton's constant and ${\cal V}_n$ is the
volume of the spherical hypersurface described by $d\Omega^2_n$.)
It is a straightforward process to generalize the
prior Painleve-de Sitter formalism for this
black hole spacetime. In particular, Eq.(\ref{6}) and Eq.(\ref{8})
should respectively be modified as follows:
\be
ds^2=-\left(1-{r^2\over l^2}-{M\epsilon_n\over r^{n-1}}
\right)d\tau^2 - 2\sqrt{{r^2\over l^2}+ {M\epsilon_n\over r^{n-1}}}
d\tau dr 
+ dr^2+r^2 d\Omega^2_n,
\label{15}
\ee
\be
{\dot r}=\sqrt{{r^2\over l^2}+ {M\epsilon_n\over r^{n-1}}} \pm 1.
\label{16}
\ee
\par
On the basis of our prior discussion, it  follows
directly that the positive-energy shell sees the effective metric
of Eq.(\ref{15}), although with $M$ replaced by
$\omega^{\prime}$. The same substitution in Eq.(\ref{16})
yields the desired expression for ${\dot r}$ as
a function of $\omega^{\prime}$. (Note that
we must choose the negative sign in Eq.(\ref{16}), as
the shell is propagating from larger to smaller $r$.) Thus, we can
now rewrite Eq.(\ref{12.5}) in the following
explicit manner: 
\be
Im\I=-Im\int^{r_f}_{r_i}\int^{\omega}_{0} {d\omega^{\prime}dr\over 
\sqrt {{r^2\over l^2}+ {\epsilon_n(\omega^{\prime}-i\delta)\over r^{n-1}}} 
- 1}.
\label{17}
\ee
Here, we have also added a small imaginary part to the effective
energy (i.e., $\omega^{\prime}\rightarrow\omega^{\prime}-i\delta$
with $\delta <<1$),
so that the above integral can be evaluated via contour techniques.
Let us further point out that $\delta>0$  is to be implied,
as this choice ensures that the positive-frequency solution
($\sim e^{-i\omega\tau}$) decays exponentially in time. 
\par
To explicitly evaluate this integral, let
us temporarily  treat $r$ as a constant and  make the following change of
variables: $x=\sqrt{{r^2\over l^2}+ {\epsilon_n\omega^{\prime}\over r^{n-1}}}$.
This  leads to:
\be
Im\I=-{2\over \epsilon_n}Im\int^{r_f}_{r_i}r^{n-1}dr
\int^{x(\omega)}_{x(0)} {x dx\over 
x - 1-i{\overline \delta}(r)},
\label{18}
\ee
where ${\overline\delta}(r)=l\epsilon_n\delta/2r^{n}\approx 0^{+}$.
Given that $x$ monotonically increases with $\omega^{\prime}$
and ${\overline\delta}>0$, it is appropriate to integrate
in a counter-clockwise direction in the upper half
of the complex-$x$ plane. Following this prescription,
we obtain:
\be
Im\I=-{2\pi\over\epsilon_n}Im i\int^{r_f}_{r_i}r^{n-1}dr.
\label{19}
\ee
\par
The integration over $r$ can now be trivially performed to give:
\be
Im\I={2\pi\over n\epsilon_n}\left[r_i^n-r_f^n\right].
\label{20}
\ee
Note that, by construction, $r_i>r_f$,  and so
the sign of $Im\I$ comes out positive as
required; cf. Eqs.(\ref{9},\ref{10}).
Generating the correct sign in de Sitter thermodynamics
is not  as trivial as one may think. Indeed, naive application
of the first law of thermodynamics to a cosmological horizon
can often lead to an  erroneous 
negative sign \cite{SSV,MYU}.
\par
The above formula is the key quantitative result of
this paper. We can  substantiate its validity  by
considering a specific value of $n$. It is readily
shown that the case of $n=1$ (i.e., 3-dimensional de Sitter
space) is in agreement with the analogous expression found
in \cite{PAR}. Another convenient choice is $n=3$
(i.e., 5-dimensional de Sitter space), 
as the Schwarzschild-de Sitter horizon can then be solved 
 for via a quadratic relation.
\par
With our attention on the $n=3$ case, it follows
(cf. Eq.(\ref{13})) that the Schwarzschild-de Sitter cosmological horizon
is described by the largest root of:
\be
{r^4_H\over l^2}-r_H^2+M\epsilon_3=0.
\label{21}
\ee
That is:
\be
r_H^2(M)={l^2\over 2}\left[1+\sqrt{1-{4M\epsilon_3\over l^2}}\right].
\label{22}
\ee
\par
Recalling our prior definitions of $r_i$ and $r_f$,
we have $r_i^2=r_H^2(0)=l^2$ and $r_f^2=r_H^2(\omega)$.
Hence, Eq.(\ref{20}) can be re-expressed  as:
\be
Im\I={2\pi l^3\over 3\epsilon_3}\left[1-{1\over 2^{3/2}}
\left(1+\sqrt{1-{4\omega\epsilon_3\over l^2}}\right)^{3/2}
\right].
\label{23}
\ee
When the particle energy is small (i.e., $\omega<<l^2/\epsilon_3$),
the above expression can be expanded to yield:
\be
Im\I=\pi l \omega+{\cal O}(\omega^2).
\label{24}
\ee
\par
Incorporating the above expansion into Eq.(\ref{10}), we are able to deduce
the temperature of radiation:
\be
T(\omega)\approx{1\over 2\pi l}+{\cal O}(\omega).
\label{25}
\ee
Reassuringly, the leading-order, energy-independent term is the well-known
background temperature of empty de Sitter space \cite{GH}.
Meanwhile, the energy-dependent corrections, which can
easily  be computed to any desired order in $\omega$,
are indicative of a ``greybody'' factor in the emission spectrum;
that is, a deviation from pure thermality. That such deviations
occur for Hawking-like radiation is well known \cite{HAW},
but this point is rarely stressed
in the relevant literature.
\par
As a further check on our formalism (this time for any $n$), 
we can consider the change in entropy
during the process of emission. The first law of thermodynamics
indicates that:
\be
\Delta S= -{\omega\over T}\approx -2Im \I = {4\pi\over n\epsilon_n}
\left[r^n_f-r^n_i\right],
\label{26}
\ee
where we have also applied Eqs.(\ref{10},\ref{20}).
We can compare this outcome with that predicted by
the Bekenstein-Hawking ``area'' law \cite{BEK,HAW2},
which tells us:
\be
\Delta S= S_f-S_i= {{\cal V}_n\over 4G_{n+2}}\left[r_f^n-r_i^n\right].
\label{29}
\ee
Since $\epsilon_{n}=16\pi G_{n+2}/n{\cal V}_n$, these two independent 
formulations of $\Delta S$ are, indeed, in perfect agreement.

\section{Schwarzschild-de Sitter Tunneling}

In the discussion to follow, we will  consider
the implications (on the tunneling picture) 
when the initial state is described by a Schwarzschild-de Sitter spacetime.  
Let us, once again, consider
the incoming radiation from the cosmological
horizon and, for the moment, ignore the outgoing
radiation from the black hole horizon.
It is readily  observed that the key result of last section, Eq.(\ref{20}),
remains valid; although  $r_i$ and $r_f$  must 
be  appropriately redefined.  Recalling the 
inverse correspondence between black hole mass and
background energy, we expect 
a black hole  of initial mass $M$ 
to  have a final mass of  $M+\omega$ (where, as before, $\omega$
is the energy of the emitted particle). It follows (cf. Eq.(\ref{13}))
that the radii in question correspond to the largest
roots of: 
\be
{r_i^2\over l^2}-1+{\epsilon_n M\over r_i^{n-1}}=0,
\label{30}
\ee
\be
{r_f^2\over l^2}-1+{\epsilon_n (M+\omega)\over r_f^{n-1}}=0.
\label{31}
\ee
\par
As in the preceding section, let us turn to the case
of $n=3$ as a check on our formalism. In this 5-dimensional
case, the above equations can be explicitly solved to yield:
\be
r_i^2={l^2\over 2}\left[1+\sqrt{1-{4\epsilon_3 M\over l^2}}\right],
\label{32}
\ee
\be
r_f^2={l^2\over 2}\left[1+\sqrt{1-{4\epsilon_3 (M+\omega)\over l^2}}\right].
\label{33}
\ee
Substituting these expressions into Eq.(\ref{20}) and  expanding,
we  find:
\be
Im\I={\pi l^2  r_i\over 2r_i^2-l^2}\omega +{\cal O}(\omega^2).
\label{34}
\ee
\par 
From the above result and Eq.(\ref{10}), the corresponding temperature 
 is found to be:
\be
T\approx{2r_i^2-l^2\over 2\pi r_i  l^2} +{\cal O}(\omega).
\label{35}
\ee
It is not difficult to verify that the leading-order term
agrees with the usual Hawking definition \cite{HAW}
(translated to a cosmological horizon \cite{GH}); that is:
\be
T={1\over 4\pi} \left|{d\over dr}\left[1-{r^2\over l^2}-{\epsilon_3 M
\over r^2}\right]\right|_{r=r_i}.
\label{36}
\ee
Furthermore, the change in entropy during emission 
can again be shown to agree with that predicted
by the Bekenstein-Hawking area law.  (See the end of Section 2
for details.)
\par
There is an intriguing  observation that follows
from the emission rate, $\Gamma\approx e^{-2Im\I}$,
being a measurable and, hence, real quantity.
Again focusing on the case of $n=3$ (although the discussion 
throughout this section
is quite general\footnote{The  generality of this discussion 
 does not, however, necessarily  apply to the $n=1$ case.
This is because the 3-dimensional Schwarzschild-de Sitter solution
describes a conical deficit angle rather than a black hole \cite{SSV}.
For  related discussion that highlights  this 3-dimensional scenario, see 
\cite{PAR}.}), we can see from  Eqs.(\ref{20},\ref{32},\ref{33}) 
 that the condition:
\be
M+\omega\leq {l^2\over 4\epsilon_3}= {3\pi l^2\over 32 G_5}
\label{37}
\ee
must always be enforced.  It is of interest that this upper bound   
corresponds precisely with the mass of the (5-dimensional) 
Nariai black hole \cite{NAR}.
Significantly, the Nariai solution describes the coincidence of the
black hole and cosmological horizons (the black hole horizon is
located by changing the explicit $+$ in Eq.(\ref{32})  
to a $-$); meaning that this solution   represents
the most massive black hole  in an asymptotically de Sitter
spacetime. Hence, the tunneling formalism provides a
natural mechanism for censoring against larger
values of mass.  Similar observations have
been made with regard to charged (Reissner-Nordstrom) black holes, where
the tunneling formalism has been shown to censor against naked 
singularities \cite{KW2,PW}.
\par
The overall picture for Schwarzschild-de Sitter space
is, however, much more complicated than we have alluded
to above. This is because radiation is both propagating inwards from the
cosmological horizon and outwards from the black hole horizon.
A formal, complete analysis must consider both of these
effects, and there would undoubtedly be 
scattering taking place between the black hole and cosmological 
contributions. Even  without delving into calculational
specifics, we can still comment on the stability
of the total  system. Once again turning to  our
$n=3$ chestnut, let us take note of the following 
(lowest-order) expressions for the temperature
(associated with the cosmological and black hole horizon, 
respectively):
\be
T_{CH}={\sqrt{1-{4\epsilon_3M\over l^2}}\over
\sqrt{2}\pi l\left[1+\sqrt{1-{4\epsilon_3M\over l^2}}\right]^{1/2}},
\label{38}
\ee
\be
T_{BH}={\sqrt{1-{4\epsilon_3M\over l^2}}\over
\sqrt{2}\pi l\left[1-\sqrt{1-{4\epsilon_3M\over l^2}}\right]^{1/2}}.
\label{39}
\ee
Here, we have applied Eqs.(\ref{36} and \ref{32})
and again note that the black hole horizon can
be found by reversing the explicit $+$ sign in Eq.(\ref{32}).  
\par
With an inspection of the above, it becomes evident that 
$T_{CH}\leq T_{BH}$; with saturation occurring 
only at the Nariai value of mass ($M=l^2/4\epsilon_3$),
in which case  both temperatures are vanishing.
With this observation, we are able to
deduce that the net flow of radiation
will always be (up to insignificant quantum fluctuations)
towards the cosmological horizon. That is to say,
the system will inevitably evolve towards  empty
de Sitter space. This phenomena is supported
by the second law of thermodynamics, since the
total entropy of a Schwarzschild-de Sitter spacetime
(or virtually any ``well-behaved''\footnote{In this context,
well-behaved implies no naked singularities and
matter that satisfies the standard energy conditions \cite{WALD}.}
 asymptotically de Sitter spacetime) is known to be bounded from
above by the entropy of empty de Sitter space \cite{BOU}.
\par
The above viewpoint  can also be substantiated 
by way of holographic (or dS/CFT duality) considerations.
In particular,  let us take note of Strominger's
realization \cite{STR2} (also see \cite{BDM,HAL}) 
that time evolution in an asymptotically
de Sitter spacetime is dual to an {\it inverted} renormalization
group flow.\footnote{We remind the reader that the renormalization group
is normally regarded as flowing from the ultraviolet (relatively large
number of degrees of freedom) to the infrared (relatively
small number of degrees of freedom).} 
On this basis, it follows that degrees of freedom
will be integrated into the system  with forward evolution
 in time. Moreover, the maximal entropic state (i.e., empty
de Sitter space) will naturally  correspond with a stable, ultraviolet fixed
point for the flow.
\par
On the other hand, because of the vanishing temperature
associated with the Nariai solution, one might  expect
the system to stabilize precisely when the horizons coincide.
Such stability would indeed be feasible 
at a strictly classical level; however, once quantum (or semi-classical)
effects are accounted for, it becomes evident that the Nariai solution
is unstable under the smallest of perturbations (see \cite{BOU2}
and references within). In renormalization group language, this Nariai
solution can be identified with  an  infrared
fixed point that is unstable \cite{HAL}.

\section{``The Dark Side of the Moon''}
\par

In this section, we will investigate  the following question: how  
would the semi-classical
 tunneling picture  be perceived by a hypothetical observer
who is trapped  outside of the cosmological horizon?
Such a query may appear to be  of little relevance,
given that a ``standard'' de Sitter observer
is causally restricted to the interior of his/her horizon.
However, here we will argue that this question merits
consideration on the basis of  dS/CFT holography.
\par
The dS/CFT duality, as we currently understand it, incorporates
the entire spacetime into its framework and not
just the causal diamond. Indeed, the dually related
conformal field theory has been conjectured  
to ``live''   on   the spacelike asymptotic boundaries \cite{STR}; these
being  future (${\cal I}^+$) and past  (${\cal I}^-$)
infinity. Significantly, both of these boundaries
lie outside of an observer's  causal diamond;
in fact, an observer  can only access precisely
one point  at either infinity.
Moreover, the only measurable (gauge-invariant)
quantities in de Sitter space would appear to be the
elements of an $S$-like matrix \cite{WIT2}
that can be expressed in terms of 
correlation functions of the dual boundary theory \cite{MSB,SV}.
To make operational sense of such ``meta-observables'' \cite{WIT2}
clearly  requires a ``special'' observer with a global view of the
entire spacetime. 
To put it another way, if a quantum theory of de Sitter gravity
is to be realized,
  we may yet have to adapt
our intuitive  ideas of what constitutes a physical
observable.
\par
With the above discussion in mind, let us return to
the quantum-tunneling description of de Sitter
radiation, as elaborated on in Section 2.
From the perspective of someone (or something?)
outside of the horizon, a negative-energy shell
is tunneling outwards.  Meanwhile, the positive-energy
partner remains behind (i.e., in the vicinity
of the horizon) and effectively raises the
energy of the background spacetime.
Hence, the effective metric, as seen by this  negative-energy
shell, must be one in which the background energy increases
with increasing   $|\omega^{\prime}|$  (i.e., the magnitude
of the shell energy, which increases from $0$ to $|\omega|$). 
We can obtain just such an effective geometry by replacing 
  $M$ with  $-|\omega^{\prime}|$  
in  the Schwarzschild-de Sitter metric of Eq.(\ref{13}). 
That is:
\be
ds^2=-\left(1-{r^2\over l^2}
+{\epsilon_n|\omega^{\prime}|\over r^{n-1}}\right)dt^2+
      \left(1-{r^2\over l^2}+{\epsilon_n|\omega^{\prime}|\over r^{n-1}}
\right)^{-1}dr^2+r^2d\Omega_n^2.
\label{40}
\ee
\par
The above metric can  readily be identified with that of the
so-called ``topological'' de Sitter
spacetime\footnote{In its most general form, the topological
de Sitter solution can allow for a hyperbolic, flat, or
(as depicted above) a spherical horizon
geometry. To obtain the hyperbolic (flat) topological solution, 
one can replace $1$ with $-1$ ($0$)  in the lapse function of Eq.(\ref{40}).}
 \cite{CMZ,CAI}. (Also see \cite{MCI} for
a recent discussion and  references.) 
From a dS/CFT perspective, the topological de Sitter
solution has the desirable property of an (apparently) unitary  boundary
 \cite{CAI,MED2}. (Conversely, the conventional Schwarzschild-de Sitter
solution would appear to have a non-unitary dual \cite{STR,CAI}.)
On the other hand, topological de Sitter spacetimes
have the detrimental feature of a naked
singularity, as there is no longer a black hole horizon
(although the cosmological horizon remains intact).
The need to universally censor against such a singularity can, however, be 
debated. That is to say, an observer outside of the
cosmological horizon would be causally disconnected from the singularity 
and need not be aware of its existence.\footnote{Although
the topological de Sitter solution has recently 
been the subject of further  criticism 
(based on  string-theoretical considerations) \cite{MCI}, this analysis
specifically applied  to  a hyperbolic horizon
geometry and is not of issue in the current discussion.}  
\par
To obtain an ``outside-of-the-horizon'' emission rate,
we can  essentially repeat the calculations of Section 2,
except using Eq.(\ref{40}) for the  effective metric and a few trivial 
 modifications.\footnote{Specifically, the radial motion
is now  outgoing
so that ${\dot r}=\sqrt{{r^2\over l^2}-{\epsilon_n |\omega^{\prime}|
\over r^{n-1}}}+1$,
 and $dH\sim +d|\omega^{\prime}|$ since  the background energy
 increases with increasing $|\omega^{\prime}|$.} 
Keeping a very careful track of the signs, we find the imaginary
part of the action to be as follows: 
\be
Im\I={2\pi\over n\epsilon_n}\left[r_f^n-r_i^n\right].
\label{41}
\ee
That is, the negative of the prior result (\ref{20}).
However, this   sign reversal
is a most welcome outcome, as now  
we have that $r_f > r_i$. (This must be the case by construction. 
It can also be  verified with an  explicit 
calculation of the horizon position as a function of particle energy. 
For instance, for $n=3$,
one finds that 
$r_i^2=l^2$ and 
$r_f^2={l^2\over 2}\left[1
+\sqrt{1+{4\epsilon_3 |\omega|\over l^2}}\right]$.)
The positivity of Eq.(\ref{41}) tells us that the effective temperature  
is strictly non-negative,
even outside of the horizon, as is  necessary for a sensible
 interpretation of the tunneling phenomena.
\par
What (if anything) have we learned from  this section?
At the risk of straying from physics to philosophy,
we propose the following pair of conjectural points: \\
{\it (i)} The topological de Sitter  geometry should not be regarded
 as a substitute for its Schwarzschild-de Sitter counterpart
but, rather, as a complementary description. 
 The choice one should make depends
on the side of the horizon under consideration.   \\
{\it (ii)} The topological de Sitter solution 
 is a
necessary ingredient if one is to
take a global view of de Sitter space. Let us re-emphasize
that such a  view is implicitly advocated
by  the dS/CFT correspondence.

\section{Conclusion}
\par

In the preceding paper, we have  considered
 de Sitter radiation as a semi-classical tunneling process.
 Adapting  the methodology of Kraus, Wilczek \cite{KW} and
others (including  a recent, related work by Parikh \cite{PAR}), 
we were able to calculate the  rate
of particle emission 
from a cosmological horizon.   
We then verified that this calculation
agreed, up to higher-order corrections, with the known
  thermodynamic properties
of de Sitter space, as well as Schwarzschild-de Sitter
space. Meanwhile, these frequency-dependent corrections indicate
that the emission spectrum of Hawking-like radiation deviates from 
perfect thermality;
a well-known but often forgotten result \cite{HAW}.
\par
Along the way, we have also touched base with certain aspects
of the dS/CFT holographic correspondence \cite{STR}.
It is quite possible that there  are  deep
connections between  semi-classical thermodynamics
and de Sitter holography that await to be uncovered.  
We hope to report progress along these lines
at a future date.

\section{Acknowledgments}
\par
The author  would like to thank  V.P.  Frolov  for helpful
conversations. 
  \par\vspace*{20pt}


\end{document}